	\documentclass[pre,twocolumn,aps]{revtex4-1}
	\usepackage{amsmath,bm}
	\usepackage{graphicx}
	\usepackage{amssymb}
	\usepackage{color}
	\usepackage{lipsum}
	\usepackage[normalem]{ulem}

	\newcommand*\diff{\mathop{}\!\mathrm{d}}

	\newcommand{\beq}{\begin{equation}}
	\newcommand{\eeq}[1]{\label{#1}\end{equation}}
	\newcommand{\brk}[1]{\left(#1\right)}

	\newcommand{\figref}[1]{Fig.~\ref{#1}}

	\newcommand{\g}{\mathbf{g}}
	
	\newcommand{\gbar}{\bar{\mathbf{g}}}
	\newcommand{\gbarA}{\bar{\mathbf{g}}_\text{A}}
	\newcommand{\gbarP}{\bar{\mathbf{g}}_\text{P}}
	
	\newcommand{\gbarO}{\bar{\mathbf{g}}^0}

	\newcommand{\A}{\mathcal{A}}

	\newcommand{\AO}{\mathcal{A}_\text{0}}

	\newcommand{\xvec}{\mathbf{x}}

	\newcommand{\T}[1]{\text{#1}}

	\newcommand{\mcm}[1]{\textcolor{black}{#1}}
	\newcommand{\mjb}[1]{\textcolor{black}{#1}}

\begin{document}
		\title{Geometric frustration and \mcm{solid-solid} transitions in model 2D tissue}
		\author{Michael Moshe$^{1,2}$,  Mark J. Bowick$^{2,3}$ and M. Cristina Marchetti$^{2}$}
		\affiliation{$^1$Department of Physics, Harvard University, Cambridge, MA 02138, USA}
		\affiliation{$^2$Department of Physics and Soft Matter Program, Syracuse University, Syracuse, NY, 13244}
		\affiliation{$^3$Kavli Institute for Theoretical Physics, University of California, Santa Barbara, CA 93106, USA}		
		\begin{abstract}
			We study the mechanical behavior of two-dimensional cellular tissues by formulating the continuum limit of discrete vertex models based on an energy that penalizes departures from a target area $A_0$ and a target perimeter $P_0$ for the component cells of the tissue. As the dimensionless {target shape index} $s_0 = \frac{P_0}{\sqrt{A_0}}$ is varied, we find a transition from a soft elastic regime for compatible target perimeter and area to a stiffer nonlinear elastic regime frustrated by geometric incompatibility. We show that the ground state in the soft regime has a family of degenerate solutions associated with zero modes for the target area and perimeter. The onset of geometric incompatibility at a critical $s_0^c$ lifts this degeneracy. The resultant energy gap leads to a nonlinear elastic response distinct from that obtained in classical elasticity models. We draw an analogy between cellular tissues and anelastic deformations in solids.
			
		\end{abstract}
		\maketitle

		Living tissues are far-from-equilibrium materials capable of spontaneously undergoing large-scale remodeling and adapting their mechanical behavior in response to internal and external cues. The experimental observation of glassy dynamics in epithelia \cite{Angelini2010,Angelini2011,Nnetu2012} has  motivated \mcm{interest in quantifying the relation between rheological and structural properties of tissue.} The aim is to provide a framework for organizing biological data by describing tissue as a material, with mechanical behavior tuned by effective parameters that provide a coarse-grained description of both intra- and inter-cellular interactions.  Significant progress has been made  \mcm{using a class of models 
that describe a confluent cell monolayer, in which there are no gaps or overlaps between cells, as a tiling of space.}
\mcm{Each cell is a polygon (Vertex Models - VM)~\cite{Honda1978, Farhadifar2007, Hufnagel2007,Nagai2001,Staple2010} or a Voronoi area (Voronoi Models)~\cite{Bi2016,su2016}, with the polygons' vertices or the Voronoi centers taken as the degrees of freedom. \mcm{ For uniform cell edge tensions,} both models are} based on a tissue energy that \mjb{penalizes} deviations of the cell area \mcm{$A$} and perimeter \mcm{$P$} from prescribed target values $A_0$ and $P_0$. Numerical solutions of \mcm{Vertex and Voronoi Models  with \textit{disordered} polygonal configurations}  \mjb{reveal}  rich behavior. \mcm{Most interesting is} the prediction of a rigidity transition tuned by cellular shape, as quantified by the dimensionless target cell shape index  $s_0=P_0/\sqrt{A_0}$, which is in turn controlled by cell-cell adhesion and cortex contractility  \cite{Bi2014,Bi2015,Bi2016,Barton2017,su2016}. 
		
		A powerful tool for describing the mechanical properties of matter is continuum elasticity. While continuum models of epithelia have been developed and used to describe biological processes, such as wound healing and morphogenesis \cite{Banerjee2015,Kopf2013,Ranft2010}, the development of a continuum theory that incorporates the rich behavior of \mcm{the tissue energy used in discrete Vertex and Voronoi models} remains an open challenge~\footnote{From here on we use the term Vertex Models (VMs) to refer to discrete models with the tissue energy given below in Eq. (1).}. Here we tackle this challenge by considering a regular polygonal tiling and develop a geometric formulation of \mcm{VM} elasticity similar to that used to describe disordered solids or materials that exhibit non-uniform differential growth, such as plant leaves \cite{Sharon2004,Klein2007,Armon2011,Kupferman2013ARMA,Moshe2016}. \mcm{ Our model is a coarse-grained version of a Vertex model, albeit with uniform cell edge tensions, because it allows for shape changes through variations in the position of the vertices of the polygons.}
{At a critical  $s_0$ ($s_0^c$), corresponding to the isoperimetric quotient  \cite{Isoperimetric}, we find a transition between a soft and a stiff solid. In the soft solid ($s_0 > s_0^c$) the target area and perimeter of individual cells are simultaneously satisfiable (compatible), whereas in the stiff solid ($s_0 < s_0^c$) they are not (incompatible).}   This geometric frustration is associated with a sharp rise in the effective stiffness of the tissue and the onset of residual stresses. The critical value of the {target} shape index depends on the  geometry of the unit cell, with $s_0^c=\sqrt{8\sqrt{3}}\simeq 3.722$ for hexagons. 
For $s_0 < s_0^c$ no hexagonal polygon exists, \mjb{supporting}  a geometric origin for the transition. 
	Earlier work  has shown that the hexagonal ground state of the \mcm{VM} is linearly unstable for $s_0>s_0^c$, where it is replaced by a soft   network of irregular polygons~\cite{Staple2010}. Here we show that
	the ground state of the soft solid is actually a family of degenerate area-preserving and perimeter-preserving states with a band of zero modes. The onset of geometric incompatibility below $s_0^c$ lifts this degeneracy and results in an energy gap, leading to a 
	finite residual stress or prestress, as commonly seen in living tissues \cite{Kasza2007}. Although our starting point is a continuum elastic energy quadratic in the strain, which would suggest a linear response at all values of imposed deformations, the force extension curves are nonlinear due to the degeneracy of the target configurations.  \mcm{The incompatible tissue shows strain stiffening, which is observed ubiquitously in living cells~ \cite{Levental2007,Fernandez2006}.} 
	
	\mcm{The solid-solid (SS) stiffening transition obtained here for regular polygonal tilings is distinct from the solid-liquid (SL) rigidity transition predicted earlier for disordered tilings~\cite{Bi2014,Bi2015}. The latter is driven by the vanishing of the energy barriers for $T1$ transformations, which are forbidden in our model. Our work suggests that the SL rigidity transition  may be facilitated by a SS  transition, where the \mcm{effective} Young's modulus of the soft solid phase becomes very small, thus easing $T1$ transformations. This is supported by the recent suggestion that the SL rigidity transition} in a disordered Voronoi model of $3D$ cellular agglomerates \mcm{may also be associated with underlying geometric incompatibility~\cite {Merkel2017}.} The geometric formulation of elasticity used here \mcm{ highlights the underlying geometric nature of SS and SL transitions in Vertex and Voronoi models}  and allows for the analytical calculation of stress-strain curves for regular lattices. The formalism can also be extended to incorporate disordered structures.
		
	
	 In \mcm{VMs}, cells are modeled as polygons that
	 can independently adjust their area $A_i$ and perimeter $P_i$ according to the energy \cite{Honda1978, Nagai2001, Farhadifar2007,Staple2010}
		\begin{equation}
	E_\text{T}=\frac{1}{2}\sum_{i}\left[\kappa_A\brk{\delta A_i/A_0}^2 A_0 +\kappa_P\brk{\delta P_i/P_0}^2 P_0  \right]\;,
	\label{eq:tissue}
\end{equation}
with 
$\delta A_i = A_i - A_0$ and $\delta P_i = P_i - P_0$.
The stiffnesses  $\kappa_A$ and $\kappa_P$  have dimensions of energy per unit area and perimeter, respectively. The first term in Eq.~\eqref{eq:tissue} arises from bulk elasticity as well as the ability of cells to adjust their area by changing their \mcm{height}. The second term describes the interplay of cortical tension and cell-cell adhesion. 
It has been shown numerically that  the disordered Vertex Model  exhibits a solid-liquid transition at  $s_0^* \approx  3.81$~\cite{Bi2014, Bi2015,Sussman2017}, where the energy barrier for bond-flipping $T1$ transitions that remodel the local cell neighborhood \mjb{vanishes}.
This prediction \mcm{for the disordered case} has been validated in bronchial cells \cite{Park2015}. A generalization  that includes cell motility has yielded a surface of solid-liquid transitions tuned by cell speed and the persistence time of single-cell dynamics \cite{Bi2016}. 
		Here we do not include $T1$ transitions, or any other topological excitations such as cell divisions. We show that even this simple limit exhibits unusual elastic behavior.

		 {\textit{Geometric formulation of tissue energy.}} In the geometric approach to linear elasticity, a thin planar sheet is described as a surface equipped with a metric, a $2 \times 2$ symmetric tensor that locally specifies the distance between points on the surface \cite{AudolyPomeauBook,Koiter1966,Efrati2009}. Simple elastic solids are characterized by a \textit{global} reference configuration, \mjb{or target metric $\mathbf{\bar{g}}$}, that is stress-free in the absence of external constraints or  loads~\footnote{For simple elastic solids $\gbar$  is Euclidean and, in Cartesian coordinates, can be written as  ${\bar{g}}_{11}={\bar{g}}_{22}=1$ and ${\bar{g}}_{12}={\bar{g}}_{21}=0$. }. The strain tensor for a deformed state with actual metric $\mathbf{g}$ is  defined  as $\mathbf{u}= \frac12(\mathbf{g}-\mathbf{\bar{g}})$. 
		The elastic energy of an isotropic Hookean solid spanning a region $\Omega$ is then given by
		\begin{equation}
		E_\T{HS} = \frac{1}{2}\int_{\Omega} \A^{\alpha\beta\gamma\delta} u_{\alpha\beta} u_{\gamma\delta} \, \diff S_{\gbar} \;,
		\label{eq:elastic}
		\end{equation}
		where $\A^{\alpha\beta\gamma\delta} =  \lambda\gbar^{\alpha\beta} \gbar^{\gamma\delta} + \mu \brk{\gbar^{\alpha\gamma} \gbar^{\beta\delta} + \gbar^{\alpha\delta} \gbar^{\beta\gamma}}$ is the elastic tensor, $\lambda$ and $\mu$ are Lam\'e constants, and $\diff S_{\gbar} = \sqrt{\det \gbar} \diff^2 \xvec$.  
		While the geometric formulation of elasticity may appear unnecessarily formal, it is  useful when describing materials laden with defects and solids with {nonuniform} differential growth that do not possess stress-free target configurations~\cite{Kroner1980}.
		In such cases the material is \textit{prestressed}, meaning that there is a residual stress even without an external load. 
		As a result there is no \textit{global} stress-free  target or reference configuration and displacement fields are consequently ill-defined. The definition of strain as a deviation of the actual metric from a target one is, however, still valid and reflects the existence of \textit{local} stress-free configurations~\cite{Efrati2009}. This formulation naturally captures the so-called \textit{incompatible elasticity} of such materials~\cite{Moshe2015PNAS, Moshe2016, Sharon2010,Klein2007}.

		Our goal is to obtain a coarse-grained form of the  {tissue}  energy of Eq.~\eqref{eq:tissue}, and to express it in terms of a local measure of strain.
		%
		Since area and perimeter can be tuned independently, different target metric tensors $\gbarA$ and $\gbarP$ are needed to characterize $A_0$ and $P_0$: \mjb{this requires} two measures of strain. A single metric tensor $\mathbf{g}$, however, characterizes the deformed state.
		Defining $\mathbf{u}^\text{A,P} = \frac{1}{2} \brk{\g - \gbar_{A,P}}$, the simplest energy functional quadratic in strains is $E = E_P + E_A $, with
		\begin{eqnarray}
		E_\text{A,P} &=& \int_{\Omega}  \frac{1}{2} {\A_\text{A,P}}^{\alpha\beta\gamma\delta} u^\text{A,P}_{\alpha\beta} u^\text{A,P}_{\gamma\delta} \diff S_{\gbar}\;,
		\label{eq:ContinuumAreaEnergy}
		\end{eqnarray}
		where
		\begin{eqnarray}
		\A_x^{\alpha\beta\gamma\delta} &= \lambda_x\gbar_x^{\alpha\beta} \gbar_x^{\gamma\delta} + \mu_x \brk{\gbar_x^{\alpha\gamma} \gbar_x^{\beta\delta} + \gbar_x^{\alpha\delta} \gbar_x^{\beta\gamma}}\;,
		\label{eq:ElasticTensors}
		\end{eqnarray}
	\mcm{with $x \in \{A,P\}$. A derivation of Eq.~\eqref{eq:ContinuumAreaEnergy} from the  discrete Eq.\eqref{eq:tissue} is carried out in the SI following the procedure used in Ref. \cite{Seung88} for flexible membranes. It confirms Eq.~\eqref{eq:ContinuumAreaEnergy} with $\lambda_P = \mu_P = 3 \kappa_P P_0/8$ and $\lambda_A = 2\mu_A = \kappa_A A_0/3$.}
	In the remainder of this work we consider for simplicity a spatially uniform system. \mcm{An example of a calculation for a tissue with nonuniform shape parameter is shown in a Mathematica notebook attached as SI.}
	
\begin{figure}
			\centerline{\includegraphics[width=0.8\columnwidth]{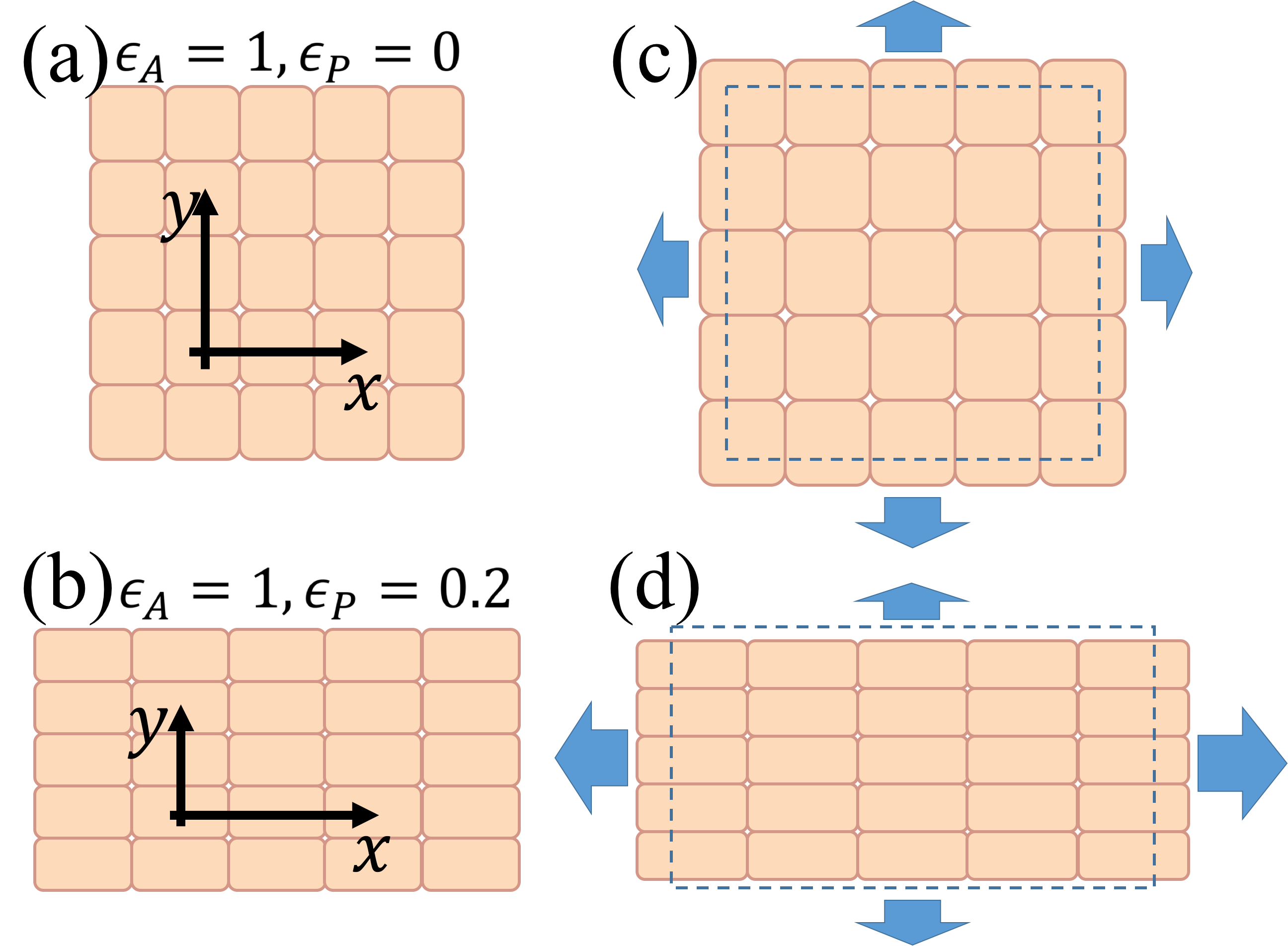}}
			\caption{ Strain measurement with respect to a family of reference configurations in a tissue that penalizes only perimeter discrepancies. (a) and (b) are two configurations with the same perimeter $P_0$; (c) and (d) are two  deformed states. Strain  should be measured by comparing each deformed configuration to the closest  reference configuration: thus (c) should be compared  to (a) and (d) compared to (b).
			}
			\label{fig:SquareTissueDemo}
		\end{figure}

		 {\textit{Ground states.} } Now comes an important subtlety {--} a given target area or perimeter can be realized by a \textit{family} of target metrics, rather than just a single one. For a lattice of  {quadrilaterals}, for instance, 
		 two families of metrics  $G_A=\{\gbarA(\epsilon_A)\}$  and $G_P=\{\gbarP(\epsilon_P)\}$ corresponding to given target area and perimeter are, respectively
		\begin{equation}
		\gbarA=
		\alpha_A\begin{bmatrix}
		\epsilon_A & 0\\
		0 & \epsilon_A^{-1}
		\end{bmatrix},
		\hspace{0.1in}
		\gbarP=
		\begin{bmatrix}
		(\alpha_P+\epsilon_P)^2 & 0\\
		0 & (\alpha_P-\epsilon_P)^2
		\end{bmatrix},
		\label{eq:metric}
		\end{equation}
		 {with $P_0=2(\sqrt{\bar{g}^P_{11}} +\sqrt{\bar{g}^P_{22}})  = 4 \alpha_{P}$ and $A_0 = \sqrt{\det \gbar_A} = \alpha_A$, both in units of the lattice constant.}
			This is  illustrated in  \figref{fig:SquareTissueDemo} for a tissue of  {quadrilateral} cells with $\kappa_A=0$.
		Frames (a) and (b) display  two ``undeformed'' configurations with $P=P_0$ that can be exchanged, \mjb{with no work,} by applying a uniaxial strain in the $x$-direction. This is commonly called a zero mode. 
		Frames (c,d) show deformed configurations with $P\not=P_0$. 
		It is evident that (c) deviates only slightly from (a) but is highly deformed compared to (b).  Similarly (d) is close to (b) but highly deformed compared to (a). {In terms of strains,} the elastic energy corresponding to small deviations of each configuration from the target one should then be calculated by comparing (c) to (a) and (d) to (b). 
		The elastic energy of a configuration of a  tissue, characterized by an actual metric $\g$,  is then given by
		%
		\begin{equation}
		E_\text{T} \brk{\g}= \min_{\gbarA \in G_A} \min_{\gbarP \in G_P} [E_\text{P} \brk{\g,\gbarP}+ E_\text{A} \brk{\g,\gbarA} ] \;.
		\label{eq:ModelEnergy}
		\end{equation}
		%
		In other words, we obtain the energy of a given a configuration $\g$  by finding the target metric that minimizes the shape energy given in Eq. \eqref{eq:ContinuumAreaEnergy}.
		In practice, we do this by minimizing the explicit expressions \mjb{given in} Eqs.~\eqref{eq:metric}
		%
		with respect to $\epsilon_A$ and $\epsilon_P$. This leads to algebraic equations, rather than the Euler-Lagrange differential equations arising from direct minimization with respect to $\g$. 
		\mcm{It is important to note that since external loads change the preferred target metrics,  the Lam\'e coefficients in \eqref{eq:ElasticTensors}  do not alone determine the elastic response of the tissue.}	
		
		%
		%

		%

	There are two classes of solutions. The first  class corresponds to configurations for which both the area and perimeter  can obtain their target values. In this case  the ground state energy  vanishes and  the tissue behaves like an anomalously soft material. The second class of solutions corresponds to the case where there are no  configurations that simultaneously satisfy the target area and perimeter, which  are \mjb{then} said to be incompatible.  The energy of the ground state \mjb{in this case} is finite. The tissue has a finite prestress  and is ``stiff'' in its response to external loads. The transition between these two classes of solutions corresponds to the stiffening of tissue. It is controlled by a purely geometric incompatibility and is independent of the specific form of the energy functional or the specific measure of strain.  
		
\begin{figure}
			\centering
			\includegraphics[width=1.0\linewidth]{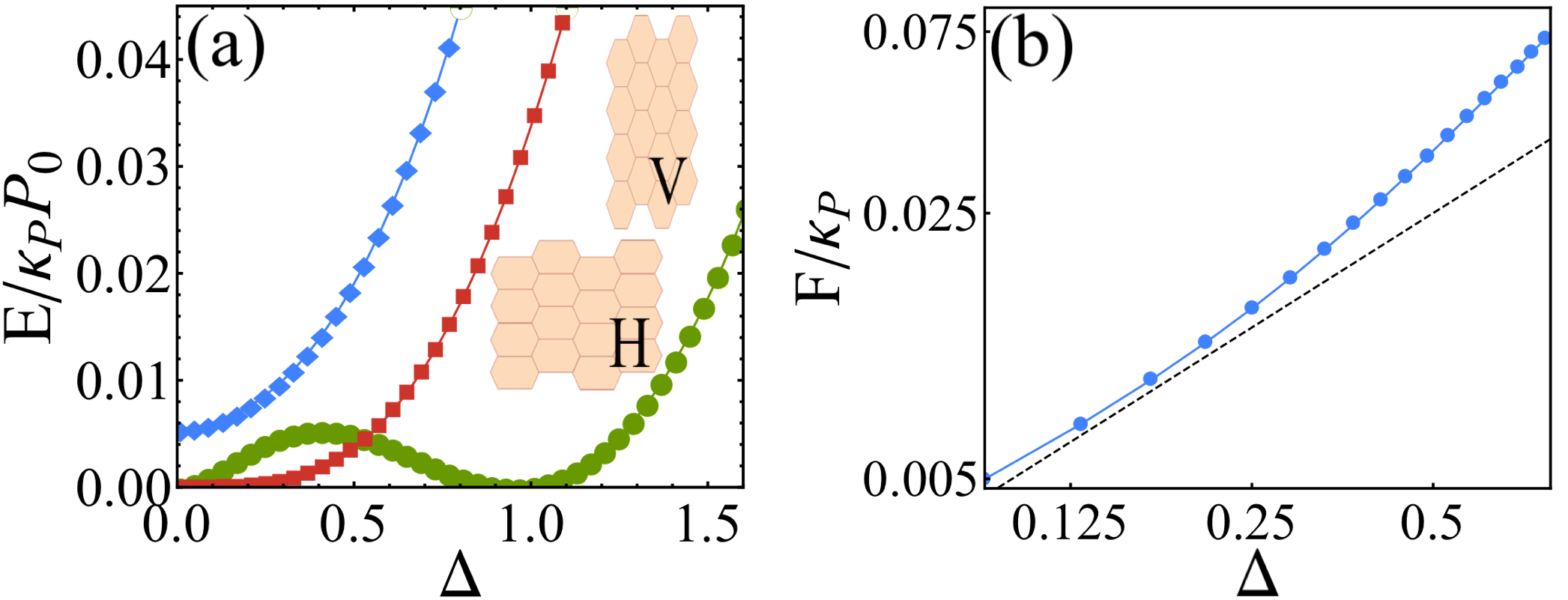}
			\caption{  (a) Dimensionless energy $E(\Delta)$  vs. strain on a linear scale  for   $s_0=3.874$ (compatible, green) , $s_0=3.722$ (at threshold, red), and $s_0=3.577$ (incompatible, blue). 
			  Two tissue configurations of equal minimal energy for $s_0=3.874$ are shown in the inset of (a).
			 The incompatible tissue has an energy gap at zero strain, corresponding to the residual stresses associated with the onset of \mjb{strain stiffening. This is  shown in (b), which displays the magnitude of the force $F=\partial E/\partial\Delta$ as a function of strain $\Delta$ on a log-log scale for $\zeta = 1$ and $s_0=3.577$. The dashed line has slope one.}
		 }
			\label{fig:energyplot}
		\end{figure}

		To demonstrate this, we now specialize to a lattice of hexagonal cells, as shown in the inset of Fig. \ref{fig:energyplot}(a). \mcm{The calculation is easily extended to other lattices of regular polygons (see SI).} To find the ground state, the symmetry of the problem allows us to neglect the off-diagonal elements of $\gbarP,\gbarA$ associated with shear zero modes.
		The parametrization of the target metrics for the area does not depend on polygonal shape: it is given by Eq.~\eqref{eq:metric}. For a hexagon whose base is oriented along the $x$ direction
		the metric $\gbarP$ is (see SI)
		\begin{equation}
		\begin{split}
		\gbar_P\brk{\epsilon_p} &= \alpha_p^2  \begin{bmatrix}
		\epsilon_p^2 & 0\\
		0 & 3-2\epsilon_p
		\end{bmatrix} \;. 
		\end{split}
		\end{equation}
%
		The unknowns characterizing the ground state are the two components  $g_{11}$ and $g_{22}$ of the actual metric and the minimizers $\epsilon_A$ and $\epsilon_P$ of the target metrics. 
		For hexagonal  cells we find that  area and perimeter are compatible for $s_0>s_0^c = \sqrt{8 \sqrt{3}}\simeq 3.722$ and the ground state energy vanishes.  For $s_0< s_0^c$ there are no compatible solutions and the ground state energy is finite, indicating that the lattice is prestressed. Note that $s_0^c$ is simply the isoperimetric value of a regular hexagon.  The nature of the solution depends only on the target shape  index $s_0$~\cite{Bi2016}, but not on $\kappa_A$ and $\kappa_P$. The ground state metric, denoted $\g^*$,  as well as the prestress, depend on the model parameters.
The solutions are always compatible if either $\kappa_A$ or $\kappa_P$ vanish. We  emphasize that while prestress is commonly associated with incompatibility between adjacent material elements, here it reflects incompatibility at the level of a single tissue element.

		%

		\textit{\mjb{Solid-Solid} transition.}
		We now consider the mechanical response of tissue to an externally applied uniaxial deformation along the $x$ direction, with the $y$ direction left free. The strain is defined as $\mathbf{u}=\g-\g^*$.
		We let $g_{11} = g^*_\text{11} + \Delta$, and determine $g_{22}(\Delta)$ by  minimizing the energy for fixed $\Delta$ (this includes minimizing with respect to $\epsilon_A$ and $\epsilon_P$).  We then evaluate the energy of this configuration to obtain $E_\text{T}(\Delta)$, shown in Fig.~\ref{fig:energyplot}(a). {Here and below energies are measured in units of $\kappa_P P_0$, and $\g$ is rescaled by $\alpha_A$, corresponding to lengths measured in units of $\sqrt{A_0}$.}
		The model is then fully characterized by two dimensionless parameters:  $s_0$ and the ratio $\zeta = \frac{\kappa_A A_0}{ \kappa_P P_0 }$ (see SI).  
		The deformation energy of the compatible tissue, displayed in  Fig. \ref{fig:energyplot}(a) (green curve), is a non-monotonic function of $\Delta$, with two degenerate minima.  This can be understood by noting that, for these parameters, the minimization described in Eq. \eqref{eq:ModelEnergy} yields the two degenerate  target configurations  shown in the inset of Fig. \ref{fig:energyplot}(a) (labelled $V$ and $H$). These can be transformed into each other via a uniaxial strain.  The energy $E(\Delta)$ shown in Fig. \ref{fig:energyplot}(a) is calculated by measuring the deformation of the lattice relative to the $V$ configuration.  When $\Delta=0$, the system is in the $V$ ground state and has zero energy. 
As $\Delta$ is increased, the lattice deforms relative to the $V$ configuration and the energy increases, eventually reaching a new zero when the deformed configuration becomes identical to the $H$ ground state. 
		The energy of the incompatible tissue (\figref{fig:energyplot}(a): \mjb{blue} curve) is also quadratic at very small $\Delta$ but has an energy gap at $\Delta=0$.
		The red curve  corresponds to the critical state at $s_0=s_0^c=3.722$. 
		In \figref{fig:energyplot}(b) we show that both compatible and incompatible tissues respond linearly at small strain, albeit with different stiffnesses. 
	
\begin{figure}
			\centering
			\includegraphics[width=1.0\linewidth]{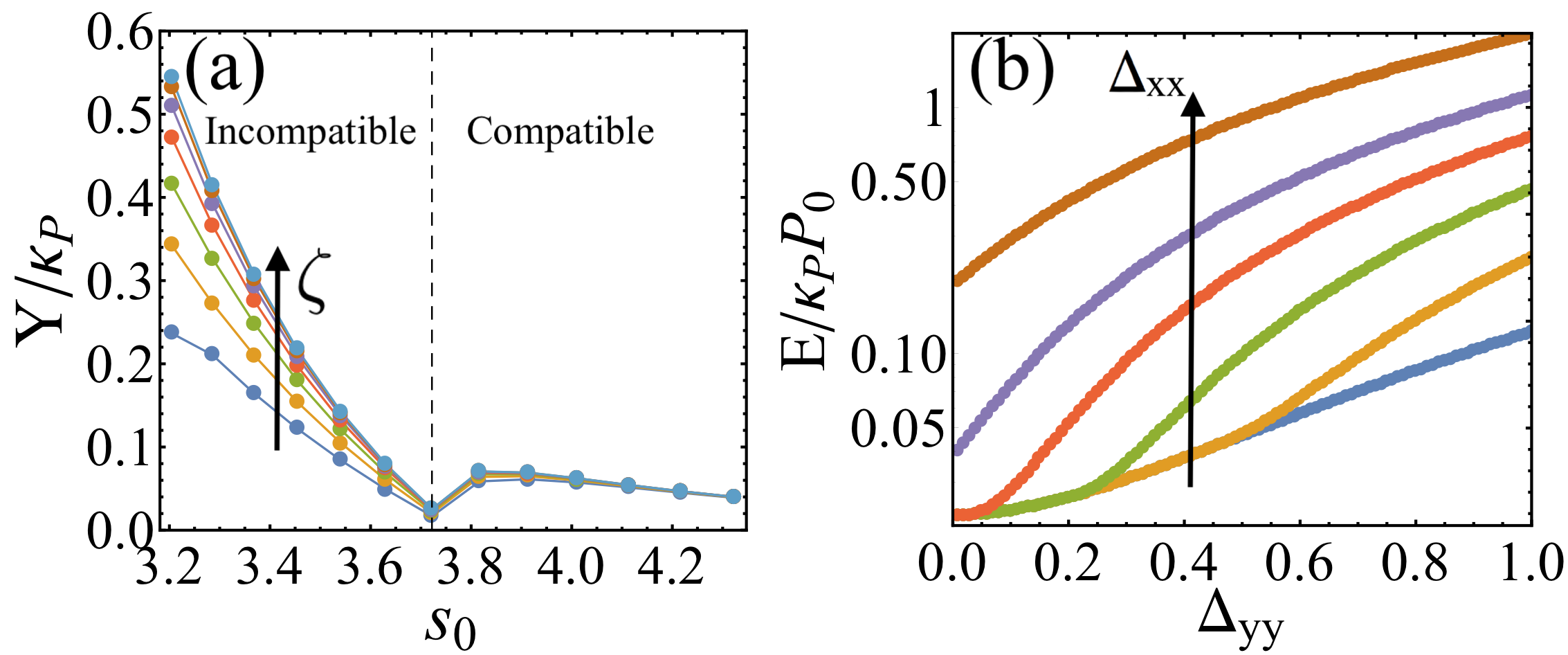}
			\caption{(a) Effective Young's modulus, \mcm{as defined by the response to stretching,} versus target shape  {index} $s_0 $ for $\zeta =0.125,0.25,0.5, 1,2,4,8$ of a uniform tissue  of hexagonal  cells.  A transition occurs at  the critical value $s_0^c = \sqrt{8 \sqrt{3}}$ (vertical dashed line) below which perimeter and area are incompatible. (b) Energy as a function of strain $\Delta_{xx}$ for various values of fixed transverse strain $\Delta_{yy} = 0.15,0.6,0.75,0.9,1.05,1.35$, for an incompatible tissue with $s_0 = 3.438$ and $\zeta = 1$.
			}
			 			
		\label{fig:figure3}
		\end{figure}

	\mjb{Tissue stiffness may be quantified by defining an effective Young's modulus $Y$ that measures the response to stretching by fitting  the energy just beyond the minimum to a quadratic form (see SI)}.  In compatible tissue $Y$ is very small for small strain, but becomes appreciable once the tissue has settled in the minimum at finite $\Delta$.  The \mcm{effective} Young's modulus is then calculated by a quadratic fit in the region beyond this second minimum. The  \mcm{effective} Young's modulus  shown in ~\figref{fig:figure3}(a) shows the onset of  stiffening at  $s_0=s_0^c$. 

		%
	The essential minimization in  \eqref{eq:ModelEnergy} renders our model tissue nonlinear at large $\Delta$.
This nonlinearity is further highlighted by noting that straining the system along a specific direction affects the mechanical response both along that direction and in the transverse direction. This is shown in 
        		\figref{fig:figure3}(b), where  we plot the elastic energy as function of a strain $\Delta_{xx}$ along the $x$ direction for various fixed strains $\Delta_{yy}$ along the $y$ direction. The corresponding effective moduli and their dependence on $\Delta_{yy}$, as well as similar figures for compatible and critically compatible tissue, are shown in SI.
		

	The connection between geometric incompatibility in cellular tissues and the emergence of \mjb{stiffness} is made \mjb{even} clearer by rewriting Eqs. \eqref{eq:ContinuumAreaEnergy} and \eqref{eq:ModelEnergy} in terms of a single effective target metric $\gbarO$. Completing the square gives
		%
		\begin{equation}
		E_\T{eff} = \min_{G_A,G_P}\frac{1}{2}\int_{\Omega} \AO^{\alpha\beta\gamma\delta} u^{0}_{\alpha\beta} u^{0}_{\gamma\delta} \sqrt{\left|\gbarO\right|}\, \T{d}^2\xvec  + E_\T{res}.
		\label{eq:effEn}
		\end{equation}
		The full expression for $\gbar^0$ in terms of $\gbar^P$ and $\gbar^A$ is given in the SI.
		The residual energy $E_\T{res}(\gbar^0,\gbar^P,\gbar^A)$  
		is independent of the actual configuration $\g$ and only depends on the families of target metrics and elastic moduli. 
		In the absence of external loads, the actual metric that minimizes the energy is $\g^* = \gbar_0$, and the optimal target metrics $\gbarA^*, \gbarP^*$, are found by minimizing the residual energy. If area and perimeter are compatible, $E_\text{res} = 0$, and the equilibrium target metrics are degenerate.
		For incompatible area and perimeter, 
		deformations  from the equilibrium target metrics are no longer zero modes, as shown by
		expanding $E_\T{res}$
		in the vicinity of the minimizers $\gbarA^* $ and $\gbarP^*$. In this case variations of the reference metrics within the families $G_A, G_P$ cost a finite energy, resulting in a gapped ground state.  \mjb{Stiffness} emerges purely as the result of this geometric frustration, as suggested by previous numerical analysis of VMs.

	\textit{Discussion.}
	Using the geometric formulation of elasticity, we have proposed a continuum energy for a two-dimensional tissue that incorporates the physics of well-established cellular tissue vertex models, \mjb{excluding T1 transformations}, and accounts for zero modes associated with area and perimeter preserving deformations.  
	We have shown that this energy yields two classes of ground states tuned by the target cell shape index $s_0$. For $s_0>s_0^c$ the tissue is soft with zero modes associated with a family of degenerate target metrics. For $s_0 {<}s_{0}^c$ one obtains a stiffer nonlinear solid with residual stress at zero external deformation \mjb{as a result of} geometric incompatibility. \mjb{An onset of stiffness} accompanied by the appearance of a residual stress was also recently demonstrated numerically in a {disordered} Voronoi model in $3D$~\cite{Merkel2017}.  Our model is purely elastic and {considers a regular lattice.}	The increase in \mjb{stiffness} is distinct from the solid-liquid (SL) transition previously reported in the literature  {for disordered tilings} and associated with the onset of finite energy barriers for $T1$ transitions. 
	\mcm{ The softening of the hexagonal lattice at $s_0=3.722$  can effectively lower the energy barriers for T1 transformations. If these are allowed, the tissue may then melt. We expect this melting to occur at $s_0=3.722$ for a regular hexagonal lattice.}
	
	\mjb{The finite energy cost of  local deformations of the target geometry resulting from incompatibilty is directly linked to the extensive literature on the plasticity of solids} (\cite{Dasgupta2012,Dasgupta2013} and references therein). \mjb{In that setting}  changes in the target geometry are interpreted as plastic deformations \cite{Efrati2013}. Fixed isotropic  inclusions in amorphous solids, \mjb{for example}, are known to strengthen the material by increasing the yield strain required for the formation of system-spanning shear bands \cite{Moshe2016}. 
%
	The analogy between variations of the reference geometry and anelastic deformations in amorphous solids suggests that introducing isotropic sources of stresses, such as inhomogeneities in the target area, may strengthen cellular tissue. This could be tested numerically. 
	
	 Finally, the formalism presented here is general and can be extended to spatially inhomogeneous target metrics to describe disordered cellular structures, or even time-varying metrics to allow for local growth. {An open question is whether the energy gap obtained here for ordered lattices will persist in disordered tilings. While a direct \mjb{extension} of our model to disordered tissue is challenging, exact and approximate analytical solutions for certain realistic problems of spatially varying shape-parameter and non-homogenous deformations are tractable, and will be presented in a future publication.}
	 \\		

		 {We thank Max Bi, Matthias Merkel and Lisa Manning for \mjb{valuable} discussions, and Yohai Bar-Sinai and Daniel Sussman for a critical reading of the manuscript.}
	We acknowledge support from the National Science Foundation at Syracuse University through DMR-1435794 (MM, MJB) and  DMR-1609208 (MCM),  at Harvard through DMR-1435999 (MM) and at KITP  through grant PHY-1125915 (MM, MJB). 
	MM acknowledges the USIEF Fulbright program.  MM and MCM acknowledge support from the Simons Foundation Targeted Grant in the Mathematical Modeling of Living Systems 342354. All authors thank the Syracuse Soft Matter Program for support and the KITP for hospitality during completion of this work.

		\bibliographystyle{apsrev4-1}
		\bibliography{references}

\begin{thebibliography}{40}%
\makeatletter
\providecommand \@ifxundefined [1]{%
 \@ifx{#1\undefined}
}%
\providecommand \@ifnum [1]{%
 \ifnum #1\expandafter \@firstoftwo
 \else \expandafter \@secondoftwo
 \fi
}%
\providecommand \@ifx [1]{%
 \ifx #1\expandafter \@firstoftwo
 \else \expandafter \@secondoftwo
 \fi
}%
\providecommand \natexlab [1]{#1}%
\providecommand \enquote  [1]{``#1''}%
\providecommand \bibnamefont  [1]{#1}%
\providecommand \bibfnamefont [1]{#1}%
\providecommand \citenamefont [1]{#1}%
\providecommand \href@noop [0]{\@secondoftwo}%
\providecommand \href [0]{\begingroup \@sanitize@url \@href}%
\providecommand \@href[1]{\@@startlink{#1}\@@href}%
\providecommand \@@href[1]{\endgroup#1\@@endlink}%
\providecommand \@sanitize@url [0]{\catcode `\\12\catcode `\$12\catcode
  `\&12\catcode `\#12\catcode `\^12\catcode `\_12\catcode `\%12\relax}%
\providecommand \@@startlink[1]{}%
\providecommand \@@endlink[0]{}%
\providecommand \url  [0]{\begingroup\@sanitize@url \@url }%
\providecommand \@url [1]{\endgroup\@href {#1}{\urlprefix }}%
\providecommand \urlprefix  [0]{URL }%
\providecommand \Eprint [0]{\href }%
\providecommand \doibase [0]{http://dx.doi.org/}%
\providecommand \selectlanguage [0]{\@gobble}%
\providecommand \bibinfo  [0]{\@secondoftwo}%
\providecommand \bibfield  [0]{\@secondoftwo}%
\providecommand \translation [1]{[#1]}%
\providecommand \BibitemOpen [0]{}%
\providecommand \bibitemStop [0]{}%
\providecommand \bibitemNoStop [0]{.\EOS\space}%
\providecommand \EOS [0]{\spacefactor3000\relax}%
\providecommand \BibitemShut  [1]{\csname bibitem#1\endcsname}%
\let\auto@bib@innerbib\@empty
\bibitem [{\citenamefont {Angelini}\ \emph {et~al.}(2010)\citenamefont
  {Angelini}, \citenamefont {Hannezo}, \citenamefont {Trepat}, \citenamefont
  {Fredberg},\ and\ \citenamefont {Weitz}}]{Angelini2010}%
  \BibitemOpen
  \bibfield  {author} {\bibinfo {author} {\bibfnamefont {T.~E.}\ \bibnamefont
  {Angelini}}, \bibinfo {author} {\bibfnamefont {E.}~\bibnamefont {Hannezo}},
  \bibinfo {author} {\bibfnamefont {X.}~\bibnamefont {Trepat}}, \bibinfo
  {author} {\bibfnamefont {J.~J.}\ \bibnamefont {Fredberg}}, \ and\ \bibinfo
  {author} {\bibfnamefont {D.~A.}\ \bibnamefont {Weitz}},\ }\href@noop {}
  {\bibfield  {journal} {\bibinfo  {journal} {Phys. Rev. Lett.}\ }\textbf
  {\bibinfo {volume} {104}},\ \bibinfo {pages} {168104} (\bibinfo {year}
  {2010})}\BibitemShut {NoStop}%
\bibitem [{\citenamefont {Angelini}\ \emph {et~al.}(2011)\citenamefont
  {Angelini}, \citenamefont {Hannezo}, \citenamefont {Trepat}, \citenamefont
  {Marquez}, \citenamefont {Fredberg},\ and\ \citenamefont
  {Weitz}}]{Angelini2011}%
  \BibitemOpen
  \bibfield  {author} {\bibinfo {author} {\bibfnamefont {T.~E.}\ \bibnamefont
  {Angelini}}, \bibinfo {author} {\bibfnamefont {E.}~\bibnamefont {Hannezo}},
  \bibinfo {author} {\bibfnamefont {X.}~\bibnamefont {Trepat}}, \bibinfo
  {author} {\bibfnamefont {M.}~\bibnamefont {Marquez}}, \bibinfo {author}
  {\bibfnamefont {J.~J.}\ \bibnamefont {Fredberg}}, \ and\ \bibinfo {author}
  {\bibfnamefont {D.~A.}\ \bibnamefont {Weitz}},\ }\href@noop {} {\bibfield
  {journal} {\bibinfo  {journal} {Proc. Natl. Acad. Sci. U.S.A}\ }\textbf
  {\bibinfo {volume} {108}},\ \bibinfo {pages} {4714} (\bibinfo {year}
  {2011})}\BibitemShut {NoStop}%
\bibitem [{\citenamefont {Nnetu}\ \emph {et~al.}(2012)\citenamefont {Nnetu},
  \citenamefont {Knorr}, \citenamefont {K{\"a}s},\ and\ \citenamefont
  {Zink}}]{Nnetu2012}%
  \BibitemOpen
  \bibfield  {author} {\bibinfo {author} {\bibfnamefont {K.~D.}\ \bibnamefont
  {Nnetu}}, \bibinfo {author} {\bibfnamefont {M.}~\bibnamefont {Knorr}},
  \bibinfo {author} {\bibfnamefont {J.}~\bibnamefont {K{\"a}s}}, \ and\
  \bibinfo {author} {\bibfnamefont {M.}~\bibnamefont {Zink}},\ }\href@noop {}
  {\bibfield  {journal} {\bibinfo  {journal} {New J Phys}\ }\textbf {\bibinfo
  {volume} {14}},\ \bibinfo {pages} {115012} (\bibinfo {year}
  {2012})}\BibitemShut {NoStop}%
\bibitem [{\citenamefont {Honda}(1978)}]{Honda1978}%
  \BibitemOpen
  \bibfield  {author} {\bibinfo {author} {\bibfnamefont {H.}~\bibnamefont
  {Honda}},\ }\href@noop {} {\bibfield  {journal} {\bibinfo  {journal} {J Theor
  Biol}\ }\textbf {\bibinfo {volume} {72}},\ \bibinfo {pages} {523IN4531}
  (\bibinfo {year} {1978})}\BibitemShut {NoStop}%
\bibitem [{\citenamefont {Farhadifar}\ \emph {et~al.}(2007)\citenamefont
  {Farhadifar}, \citenamefont {R{\"o}per}, \citenamefont {Aigouy},
  \citenamefont {Eaton},\ and\ \citenamefont {J{\"u}licher}}]{Farhadifar2007}%
  \BibitemOpen
  \bibfield  {author} {\bibinfo {author} {\bibfnamefont {R.}~\bibnamefont
  {Farhadifar}}, \bibinfo {author} {\bibfnamefont {J.-C.}\ \bibnamefont
  {R{\"o}per}}, \bibinfo {author} {\bibfnamefont {B.}~\bibnamefont {Aigouy}},
  \bibinfo {author} {\bibfnamefont {S.}~\bibnamefont {Eaton}}, \ and\ \bibinfo
  {author} {\bibfnamefont {F.}~\bibnamefont {J{\"u}licher}},\ }\href@noop {}
  {\bibfield  {journal} {\bibinfo  {journal} {Current Biology}\ }\textbf
  {\bibinfo {volume} {17}},\ \bibinfo {pages} {2095} (\bibinfo {year}
  {2007})}\BibitemShut {NoStop}%
\bibitem [{\citenamefont {Hufnagel}\ \emph {et~al.}(2007)\citenamefont
  {Hufnagel}, \citenamefont {Teleman}, \citenamefont {Rouault}, \citenamefont
  {Cohen},\ and\ \citenamefont {Shraiman}}]{Hufnagel2007}%
  \BibitemOpen
  \bibfield  {author} {\bibinfo {author} {\bibfnamefont {L.}~\bibnamefont
  {Hufnagel}}, \bibinfo {author} {\bibfnamefont {A.~A.}\ \bibnamefont
  {Teleman}}, \bibinfo {author} {\bibfnamefont {H.}~\bibnamefont {Rouault}},
  \bibinfo {author} {\bibfnamefont {S.~M.}\ \bibnamefont {Cohen}}, \ and\
  \bibinfo {author} {\bibfnamefont {B.~I.}\ \bibnamefont {Shraiman}},\
  }\href@noop {} {\bibfield  {journal} {\bibinfo  {journal} {Proc. Natl. Acad.
  Sci. U.S.A}\ }\textbf {\bibinfo {volume} {104}},\ \bibinfo {pages} {3835}
  (\bibinfo {year} {2007})}\BibitemShut {NoStop}%
\bibitem [{\citenamefont {Nagai}\ and\ \citenamefont
  {Honda}(2001)}]{Nagai2001}%
  \BibitemOpen
  \bibfield  {author} {\bibinfo {author} {\bibfnamefont {T.}~\bibnamefont
  {Nagai}}\ and\ \bibinfo {author} {\bibfnamefont {H.}~\bibnamefont {Honda}},\
  }\href@noop {} {\bibfield  {journal} {\bibinfo  {journal} {Phil. Mag. B}\
  }\textbf {\bibinfo {volume} {81}},\ \bibinfo {pages} {699} (\bibinfo {year}
  {2001})}\BibitemShut {NoStop}%
\bibitem [{\citenamefont {Staple}\ \emph {et~al.}(2010)\citenamefont {Staple},
  \citenamefont {Farhadifar}, \citenamefont {R{\"o}per}, \citenamefont
  {Aigouy}, \citenamefont {Eaton},\ and\ \citenamefont
  {J{\"u}licher}}]{Staple2010}%
  \BibitemOpen
  \bibfield  {author} {\bibinfo {author} {\bibfnamefont {D.}~\bibnamefont
  {Staple}}, \bibinfo {author} {\bibfnamefont {R.}~\bibnamefont {Farhadifar}},
  \bibinfo {author} {\bibfnamefont {J.~C.}\ \bibnamefont {R{\"o}per}}, \bibinfo
  {author} {\bibfnamefont {B.}~\bibnamefont {Aigouy}}, \bibinfo {author}
  {\bibfnamefont {S.}~\bibnamefont {Eaton}}, \ and\ \bibinfo {author}
  {\bibfnamefont {F.}~\bibnamefont {J{\"u}licher}},\ }\href@noop {} {\bibfield
  {journal} {\bibinfo  {journal} {EPJ E}\ }\textbf {\bibinfo {volume} {33}},\
  \bibinfo {pages} {117} (\bibinfo {year} {2010})}\BibitemShut {NoStop}%
\bibitem [{\citenamefont {Bi}\ \emph {et~al.}(2016)\citenamefont {Bi},
  \citenamefont {Yang}, \citenamefont {Marchetti},\ and\ \citenamefont
  {Manning}}]{Bi2016}%
  \BibitemOpen
  \bibfield  {author} {\bibinfo {author} {\bibfnamefont {D.}~\bibnamefont
  {Bi}}, \bibinfo {author} {\bibfnamefont {X.}~\bibnamefont {Yang}}, \bibinfo
  {author} {\bibfnamefont {M.~C.}\ \bibnamefont {Marchetti}}, \ and\ \bibinfo
  {author} {\bibfnamefont {M.~L.}\ \bibnamefont {Manning}},\ }\href@noop {}
  {\bibfield  {journal} {\bibinfo  {journal} {Phys. Rev. X}\ }\textbf {\bibinfo
  {volume} {6}},\ \bibinfo {pages} {021011} (\bibinfo {year}
  {2016})}\BibitemShut {NoStop}%
\bibitem [{\citenamefont {Su}\ and\ \citenamefont {Lan}(2016)}]{su2016}%
  \BibitemOpen
  \bibfield  {author} {\bibinfo {author} {\bibfnamefont {T.}~\bibnamefont
  {Su}}\ and\ \bibinfo {author} {\bibfnamefont {G.}~\bibnamefont {Lan}},\
  }\href@noop {} {\bibfield  {journal} {\bibinfo  {journal} {arXiv:1610.04254}\
  } (\bibinfo {year} {2016})}\BibitemShut {NoStop}%
\bibitem [{\citenamefont {Bi}\ \emph {et~al.}(2014)\citenamefont {Bi},
  \citenamefont {Lopez}, \citenamefont {Schwarz},\ and\ \citenamefont
  {Manning}}]{Bi2014}%
  \BibitemOpen
  \bibfield  {author} {\bibinfo {author} {\bibfnamefont {D.}~\bibnamefont
  {Bi}}, \bibinfo {author} {\bibfnamefont {J.~H.}\ \bibnamefont {Lopez}},
  \bibinfo {author} {\bibfnamefont {J.}~\bibnamefont {Schwarz}}, \ and\
  \bibinfo {author} {\bibfnamefont {M.~L.}\ \bibnamefont {Manning}},\
  }\href@noop {} {\bibfield  {journal} {\bibinfo  {journal} {Soft Matter}\
  }\textbf {\bibinfo {volume} {10}},\ \bibinfo {pages} {1885} (\bibinfo {year}
  {2014})}\BibitemShut {NoStop}%
\bibitem [{\citenamefont {Bi}\ \emph {et~al.}(2015)\citenamefont {Bi},
  \citenamefont {Lopez}, \citenamefont {Schwarz},\ and\ \citenamefont
  {Manning}}]{Bi2015}%
  \BibitemOpen
  \bibfield  {author} {\bibinfo {author} {\bibfnamefont {D.}~\bibnamefont
  {Bi}}, \bibinfo {author} {\bibfnamefont {J.}~\bibnamefont {Lopez}}, \bibinfo
  {author} {\bibfnamefont {J.}~\bibnamefont {Schwarz}}, \ and\ \bibinfo
  {author} {\bibfnamefont {M.~L.}\ \bibnamefont {Manning}},\ }\href@noop {}
  {\bibfield  {journal} {\bibinfo  {journal} {Nat. Phys.}\ }\textbf {\bibinfo
  {volume} {11}},\ \bibinfo {pages} {1074} (\bibinfo {year}
  {2015})}\BibitemShut {NoStop}%
\bibitem [{\citenamefont {Barton}\ \emph {et~al.}(2017)\citenamefont {Barton},
  \citenamefont {Henkes}, \citenamefont {Weijer},\ and\ \citenamefont
  {Sknepnek}}]{Barton2017}%
  \BibitemOpen
  \bibfield  {author} {\bibinfo {author} {\bibfnamefont {D.}~\bibnamefont
  {Barton}}, \bibinfo {author} {\bibfnamefont {S.}~\bibnamefont {Henkes}},
  \bibinfo {author} {\bibfnamefont {C.}~\bibnamefont {Weijer}}, \ and\ \bibinfo
  {author} {\bibfnamefont {R.}~\bibnamefont {Sknepnek}},\ }\href@noop {}
  {\bibfield  {journal} {\bibinfo  {journal} {PLoS Comput. Biol.}\ }\textbf
  {\bibinfo {volume} {13(6)}} (\bibinfo {year} {2017})}\BibitemShut {NoStop}%
\bibitem [{\citenamefont {Banerjee}\ \emph {et~al.}(2015)\citenamefont
  {Banerjee}, \citenamefont {Utuje},\ and\ \citenamefont
  {Marchetti}}]{Banerjee2015}%
  \BibitemOpen
  \bibfield  {author} {\bibinfo {author} {\bibfnamefont {S.}~\bibnamefont
  {Banerjee}}, \bibinfo {author} {\bibfnamefont {K.~J.~C.}\ \bibnamefont
  {Utuje}}, \ and\ \bibinfo {author} {\bibfnamefont {M.~C.}\ \bibnamefont
  {Marchetti}},\ }\href@noop {} {\bibfield  {journal} {\bibinfo  {journal}
  {Phys. Rev. Lett.}\ }\textbf {\bibinfo {volume} {114}},\ \bibinfo {pages}
  {228101} (\bibinfo {year} {2015})}\BibitemShut {NoStop}%
\bibitem [{\citenamefont {K{\"o}pf}\ and\ \citenamefont
  {Pismen}(2013)}]{Kopf2013}%
  \BibitemOpen
  \bibfield  {author} {\bibinfo {author} {\bibfnamefont {M.~H.}\ \bibnamefont
  {K{\"o}pf}}\ and\ \bibinfo {author} {\bibfnamefont {L.~M.}\ \bibnamefont
  {Pismen}},\ }\href@noop {} {\bibfield  {journal} {\bibinfo  {journal} {Soft
  Matter}\ }\textbf {\bibinfo {volume} {9}},\ \bibinfo {pages} {3727} (\bibinfo
  {year} {2013})}\BibitemShut {NoStop}%
\bibitem [{\citenamefont {Ranft}\ \emph {et~al.}(2010)\citenamefont {Ranft},
  \citenamefont {Basan}, \citenamefont {Elgeti}, \citenamefont {Joanny},
  \citenamefont {Prost},\ and\ \citenamefont {J{\"u}licher}}]{Ranft2010}%
  \BibitemOpen
  \bibfield  {author} {\bibinfo {author} {\bibfnamefont {J.}~\bibnamefont
  {Ranft}}, \bibinfo {author} {\bibfnamefont {M.}~\bibnamefont {Basan}},
  \bibinfo {author} {\bibfnamefont {J.}~\bibnamefont {Elgeti}}, \bibinfo
  {author} {\bibfnamefont {J.-F.}\ \bibnamefont {Joanny}}, \bibinfo {author}
  {\bibfnamefont {J.}~\bibnamefont {Prost}}, \ and\ \bibinfo {author}
  {\bibfnamefont {F.}~\bibnamefont {J{\"u}licher}},\ }\href@noop {} {\bibfield
  {journal} {\bibinfo  {journal} {Proc. Natl. Acad. Sci. U.S.A}\ }\textbf
  {\bibinfo {volume} {107}},\ \bibinfo {pages} {20863} (\bibinfo {year}
  {2010})}\BibitemShut {NoStop}%
\bibitem [{Note1()}]{Note1}%
  \BibitemOpen
  \bibinfo {note} {From here on we use the term Vertex Models (VMs) to refer to
  discrete models with the tissue energy given below in Eq. (1).}\BibitemShut
  {Stop}%
\bibitem [{\citenamefont {Eran}\ \emph {et~al.}(2004)\citenamefont {Eran},
  \citenamefont {Marder},\ and\ \citenamefont {Swinney}}]{Sharon2004}%
  \BibitemOpen
  \bibfield  {author} {\bibinfo {author} {\bibfnamefont {S.}~\bibnamefont
  {Eran}}, \bibinfo {author} {\bibfnamefont {M.}~\bibnamefont {Marder}}, \ and\
  \bibinfo {author} {\bibfnamefont {H.~L.}\ \bibnamefont {Swinney}},\
  }\href@noop {} {\bibfield  {journal} {\bibinfo  {journal} {American
  Scientist}\ }\textbf {\bibinfo {volume} {92}},\ \bibinfo {pages} {254}
  (\bibinfo {year} {2004})}\BibitemShut {NoStop}%
\bibitem [{\citenamefont {Klein}\ \emph {et~al.}(2007)\citenamefont {Klein},
  \citenamefont {Efrati},\ and\ \citenamefont {Sharon}}]{Klein2007}%
  \BibitemOpen
  \bibfield  {author} {\bibinfo {author} {\bibfnamefont {Y.}~\bibnamefont
  {Klein}}, \bibinfo {author} {\bibfnamefont {E.}~\bibnamefont {Efrati}}, \
  and\ \bibinfo {author} {\bibfnamefont {E.}~\bibnamefont {Sharon}},\
  }\href@noop {} {\bibfield  {journal} {\bibinfo  {journal} {Science}\ }\textbf
  {\bibinfo {volume} {315}},\ \bibinfo {pages} {1116} (\bibinfo {year}
  {2007})}\BibitemShut {NoStop}%
\bibitem [{\citenamefont {Armon}\ \emph {et~al.}(2011)\citenamefont {Armon},
  \citenamefont {Efrati}, \citenamefont {Kupferman},\ and\ \citenamefont
  {Sharon}}]{Armon2011}%
  \BibitemOpen
  \bibfield  {author} {\bibinfo {author} {\bibfnamefont {S.}~\bibnamefont
  {Armon}}, \bibinfo {author} {\bibfnamefont {E.}~\bibnamefont {Efrati}},
  \bibinfo {author} {\bibfnamefont {R.}~\bibnamefont {Kupferman}}, \ and\
  \bibinfo {author} {\bibfnamefont {E.}~\bibnamefont {Sharon}},\ }\href@noop {}
  {\bibfield  {journal} {\bibinfo  {journal} {Science}\ }\textbf {\bibinfo
  {volume} {333}},\ \bibinfo {pages} {1726} (\bibinfo {year}
  {2011})}\BibitemShut {NoStop}%
\bibitem [{\citenamefont {Kupferman}\ \emph {et~al.}(2015)\citenamefont
  {Kupferman}, \citenamefont {Moshe},\ and\ \citenamefont
  {Solomon}}]{Kupferman2013ARMA}%
  \BibitemOpen
  \bibfield  {author} {\bibinfo {author} {\bibfnamefont {R.}~\bibnamefont
  {Kupferman}}, \bibinfo {author} {\bibfnamefont {M.}~\bibnamefont {Moshe}}, \
  and\ \bibinfo {author} {\bibfnamefont {J.~P.}\ \bibnamefont {Solomon}},\
  }\href@noop {} {\bibfield  {journal} {\bibinfo  {journal} {Arch. Rat. Mech.
  Ana. , 2015}\ } (\bibinfo {year} {2015})}\BibitemShut {NoStop}%
\bibitem [{\citenamefont {Hentschel}\ \emph {et~al.}(2016)\citenamefont
  {Hentschel}, \citenamefont {Moshe}, \citenamefont {Procaccia},\ and\
  \citenamefont {Samwer}}]{Moshe2016}%
  \BibitemOpen
  \bibfield  {author} {\bibinfo {author} {\bibfnamefont {H.~G.~E.}\
  \bibnamefont {Hentschel}}, \bibinfo {author} {\bibfnamefont {M.}~\bibnamefont
  {Moshe}}, \bibinfo {author} {\bibfnamefont {I.}~\bibnamefont {Procaccia}}, \
  and\ \bibinfo {author} {\bibfnamefont {K.}~\bibnamefont {Samwer}},\
  }\href@noop {} {\bibfield  {journal} {\bibinfo  {journal} {Phil. Mag.}\
  }\textbf {\bibinfo {volume} {96}},\ \bibinfo {pages} {1399} (\bibinfo {year}
  {2016})}\BibitemShut {NoStop}%
\bibitem [{\citenamefont {Bl{\aa}sj{\"o}}(2005)}]{Isoperimetric}%
  \BibitemOpen
  \bibfield  {author} {\bibinfo {author} {\bibfnamefont {V.}~\bibnamefont
  {Bl{\aa}sj{\"o}}},\ }\href@noop {} {\bibfield  {journal} {\bibinfo  {journal}
  {Am. Math. Mon.}\ }\textbf {\bibinfo {volume} {112}},\ \bibinfo {pages} {526}
  (\bibinfo {year} {2005})}\BibitemShut {NoStop}%
\bibitem [{\citenamefont {Kasza}\ \emph {et~al.}(2007)\citenamefont {Kasza},
  \citenamefont {Rowat}, \citenamefont {Liu}, \citenamefont {Angelini},
  \citenamefont {Brangwynne}, \citenamefont {Koenderink},\ and\ \citenamefont
  {Weitz}}]{Kasza2007}%
  \BibitemOpen
  \bibfield  {author} {\bibinfo {author} {\bibfnamefont {K.~E.}\ \bibnamefont
  {Kasza}}, \bibinfo {author} {\bibfnamefont {A.~C.}\ \bibnamefont {Rowat}},
  \bibinfo {author} {\bibfnamefont {J.}~\bibnamefont {Liu}}, \bibinfo {author}
  {\bibfnamefont {T.~E.}\ \bibnamefont {Angelini}}, \bibinfo {author}
  {\bibfnamefont {C.~P.}\ \bibnamefont {Brangwynne}}, \bibinfo {author}
  {\bibfnamefont {G.~H.}\ \bibnamefont {Koenderink}}, \ and\ \bibinfo {author}
  {\bibfnamefont {D.~A.}\ \bibnamefont {Weitz}},\ }\href@noop {} {\bibfield
  {journal} {\bibinfo  {journal} {Current opinion}\ }\textbf {\bibinfo {volume}
  {19}},\ \bibinfo {pages} {101} (\bibinfo {year} {2007})}\BibitemShut
  {NoStop}%
\bibitem [{\citenamefont {Levental}\ \emph {et~al.}(2007)\citenamefont
  {Levental}, \citenamefont {Georges},\ and\ \citenamefont
  {Janmey}}]{Levental2007}%
  \BibitemOpen
  \bibfield  {author} {\bibinfo {author} {\bibfnamefont {I.}~\bibnamefont
  {Levental}}, \bibinfo {author} {\bibfnamefont {P.~C.}\ \bibnamefont
  {Georges}}, \ and\ \bibinfo {author} {\bibfnamefont {P.~A.}\ \bibnamefont
  {Janmey}},\ }\href@noop {} {\bibfield  {journal} {\bibinfo  {journal} {Soft
  Matter}\ }\textbf {\bibinfo {volume} {3}},\ \bibinfo {pages} {299} (\bibinfo
  {year} {2007})}\BibitemShut {NoStop}%
\bibitem [{\citenamefont {Fern{\'a}ndez}\ \emph {et~al.}(2006)\citenamefont
  {Fern{\'a}ndez}, \citenamefont {Pullarkat},\ and\ \citenamefont
  {Ott}}]{Fernandez2006}%
  \BibitemOpen
  \bibfield  {author} {\bibinfo {author} {\bibfnamefont {P.}~\bibnamefont
  {Fern{\'a}ndez}}, \bibinfo {author} {\bibfnamefont {P.~A.}\ \bibnamefont
  {Pullarkat}}, \ and\ \bibinfo {author} {\bibfnamefont {A.}~\bibnamefont
  {Ott}},\ }\href@noop {} {\bibfield  {journal} {\bibinfo  {journal} {Biophys.
  J}\ }\textbf {\bibinfo {volume} {90}},\ \bibinfo {pages} {3796} (\bibinfo
  {year} {2006})}\BibitemShut {NoStop}%
\bibitem [{\citenamefont {Merkel}\ and\ \citenamefont
  {Manning}(2018)}]{Merkel2017}%
  \BibitemOpen
  \bibfield  {author} {\bibinfo {author} {\bibfnamefont {M.}~\bibnamefont
  {Merkel}}\ and\ \bibinfo {author} {\bibfnamefont {M.~L.}\ \bibnamefont
  {Manning}},\ }\href@noop {} {\bibfield  {journal} {\bibinfo  {journal} {New
  J. Phys.}\ }\textbf {\bibinfo {volume} {20}},\ \bibinfo {pages} {022002}
  (\bibinfo {year} {2018})}\BibitemShut {NoStop}%
\bibitem [{\citenamefont {Sussman}\ and\ \citenamefont
  {Merkel}(2017)}]{Sussman2017}%
  \BibitemOpen
  \bibfield  {author} {\bibinfo {author} {\bibfnamefont {D.}~\bibnamefont
  {Sussman}}\ and\ \bibinfo {author} {\bibfnamefont {M.}~\bibnamefont
  {Merkel}},\ }\href@noop {} {\bibfield  {journal} {\bibinfo  {journal}
  {arxiv:1708.03396}\ } (\bibinfo {year} {2017})}\BibitemShut {NoStop}%
\bibitem [{\citenamefont {Park}\ \emph {et~al.}(2015)\citenamefont {Park},
  \citenamefont {Kim}, \citenamefont {Bi}, \citenamefont {Mitchel},
  \citenamefont {Qazvini}, \citenamefont {Tantisira}, \citenamefont {Park},
  \citenamefont {McGill}, \citenamefont {Kim}, \citenamefont {Gweon} \emph
  {et~al.}}]{Park2015}%
  \BibitemOpen
  \bibfield  {author} {\bibinfo {author} {\bibfnamefont {J.-A.}\ \bibnamefont
  {Park}}, \bibinfo {author} {\bibfnamefont {J.~H.}\ \bibnamefont {Kim}},
  \bibinfo {author} {\bibfnamefont {D.}~\bibnamefont {Bi}}, \bibinfo {author}
  {\bibfnamefont {J.~A.}\ \bibnamefont {Mitchel}}, \bibinfo {author}
  {\bibfnamefont {N.~T.}\ \bibnamefont {Qazvini}}, \bibinfo {author}
  {\bibfnamefont {K.}~\bibnamefont {Tantisira}}, \bibinfo {author}
  {\bibfnamefont {C.~Y.}\ \bibnamefont {Park}}, \bibinfo {author}
  {\bibfnamefont {M.}~\bibnamefont {McGill}}, \bibinfo {author} {\bibfnamefont
  {S.-H.}\ \bibnamefont {Kim}}, \bibinfo {author} {\bibfnamefont
  {B.}~\bibnamefont {Gweon}},  \emph {et~al.},\ }\href@noop {} {\bibfield
  {journal} {\bibinfo  {journal} {Nat Mat}\ }\textbf {\bibinfo {volume} {14}},\
  \bibinfo {pages} {1040} (\bibinfo {year} {2015})}\BibitemShut {NoStop}%
\bibitem [{\citenamefont {Audoly}\ and\ \citenamefont
  {Pomeau}(2010)}]{AudolyPomeauBook}%
  \BibitemOpen
  \bibfield  {author} {\bibinfo {author} {\bibfnamefont {B.}~\bibnamefont
  {Audoly}}\ and\ \bibinfo {author} {\bibfnamefont {Y.}~\bibnamefont
  {Pomeau}},\ }\href@noop {} {\bibfield  {journal} {\bibinfo  {journal}
  {Elasticity and geometry: from hair curls to the non-linear response of
  shells}\ } (\bibinfo {year} {2010})}\BibitemShut {NoStop}%
\bibitem [{\citenamefont {Koiter}(1966)}]{Koiter1966}%
  \BibitemOpen
  \bibfield  {author} {\bibinfo {author} {\bibfnamefont {W.~T.}\ \bibnamefont
  {Koiter}},\ }\href@noop {} {\bibfield  {journal} {\bibinfo  {journal}
  {Koninklijke Nederlandse Akademie van Wetenschappen, Proceedings, Series B}\
  }\textbf {\bibinfo {volume} {69}},\ \bibinfo {pages} {1} (\bibinfo {year}
  {1966})}\BibitemShut {NoStop}%
\bibitem [{\citenamefont {Efrati}\ \emph {et~al.}(2009)\citenamefont {Efrati},
  \citenamefont {Sharon},\ and\ \citenamefont {Kupferman}}]{Efrati2009}%
  \BibitemOpen
  \bibfield  {author} {\bibinfo {author} {\bibfnamefont {E.}~\bibnamefont
  {Efrati}}, \bibinfo {author} {\bibfnamefont {E.}~\bibnamefont {Sharon}}, \
  and\ \bibinfo {author} {\bibfnamefont {R.}~\bibnamefont {Kupferman}},\
  }\href@noop {} {\bibfield  {journal} {\bibinfo  {journal} {JMPS}\ }\textbf
  {\bibinfo {volume} {57}},\ \bibinfo {pages} {762} (\bibinfo {year}
  {2009})}\BibitemShut {NoStop}%
\bibitem [{Note2()}]{Note2}%
  \BibitemOpen
  \bibinfo {note} {For simple elastic solids $\protect \mathaccentV
  {bar}016{\protect \mathbf {g}}$ is Euclidean and, in Cartesian coordinates,
  can be written as ${\protect \mathaccentV {bar}016{g}}_{11}={\protect
  \mathaccentV {bar}016{g}}_{22}=1$ and ${\protect \mathaccentV
  {bar}016{g}}_{12}={\protect \mathaccentV {bar}016{g}}_{21}=0$.}\BibitemShut
  {Stop}%
\bibitem [{\citenamefont {Kr\"oner}(1980)}]{Kroner1980}%
  \BibitemOpen
  \bibfield  {author} {\bibinfo {author} {\bibfnamefont {E.}~\bibnamefont
  {Kr\"oner}},\ }\href@noop {} {\bibfield  {journal} {\bibinfo  {journal} {Les
  Houches}\ }\textbf {\bibinfo {volume} {35}} (\bibinfo {year}
  {1980})}\BibitemShut {NoStop}%
\bibitem [{\citenamefont {Moshe}\ \emph {et~al.}(2015)\citenamefont {Moshe},
  \citenamefont {Levin}, \citenamefont {Aharoni}, \citenamefont {Kupferman},\
  and\ \citenamefont {Sharon}}]{Moshe2015PNAS}%
  \BibitemOpen
  \bibfield  {author} {\bibinfo {author} {\bibfnamefont {M.}~\bibnamefont
  {Moshe}}, \bibinfo {author} {\bibfnamefont {I.}~\bibnamefont {Levin}},
  \bibinfo {author} {\bibfnamefont {H.}~\bibnamefont {Aharoni}}, \bibinfo
  {author} {\bibfnamefont {R.}~\bibnamefont {Kupferman}}, \ and\ \bibinfo
  {author} {\bibfnamefont {E.}~\bibnamefont {Sharon}},\ }\href@noop {}
  {\bibfield  {journal} {\bibinfo  {journal} {Proc. Natl. Acad. Sci. U.S.A}\
  }\textbf {\bibinfo {volume} {112}},\ \bibinfo {pages} {10873} (\bibinfo
  {year} {2015})}\BibitemShut {NoStop}%
\bibitem [{\citenamefont {Sharon}\ and\ \citenamefont
  {Efrati}(2010)}]{Sharon2010}%
  \BibitemOpen
  \bibfield  {author} {\bibinfo {author} {\bibfnamefont {E.}~\bibnamefont
  {Sharon}}\ and\ \bibinfo {author} {\bibfnamefont {E.}~\bibnamefont
  {Efrati}},\ }\href@noop {} {\bibfield  {journal} {\bibinfo  {journal} {Soft
  Matter}\ }\textbf {\bibinfo {volume} {6}},\ \bibinfo {pages} {5693} (\bibinfo
  {year} {2010})}\BibitemShut {NoStop}%
\bibitem [{\citenamefont {Seung}\ and\ \citenamefont {Nelson}(1988)}]{Seung88}%
  \BibitemOpen
  \bibfield  {author} {\bibinfo {author} {\bibfnamefont {S.~H.}\ \bibnamefont
  {Seung}}\ and\ \bibinfo {author} {\bibfnamefont {D.~R.}\ \bibnamefont
  {Nelson}},\ }\href@noop {} {\bibfield  {journal} {\bibinfo  {journal} {Phys.
  Rev. A}\ }\textbf {\bibinfo {volume} {38}},\ \bibinfo {pages} {1005}
  (\bibinfo {year} {1988})}\BibitemShut {NoStop}%
\bibitem [{\citenamefont {Dasgupta}\ \emph {et~al.}(2012)\citenamefont
  {Dasgupta}, \citenamefont {Hentschel},\ and\ \citenamefont
  {Procaccia}}]{Dasgupta2012}%
  \BibitemOpen
  \bibfield  {author} {\bibinfo {author} {\bibfnamefont {R.}~\bibnamefont
  {Dasgupta}}, \bibinfo {author} {\bibfnamefont {H.~G.~E.}\ \bibnamefont
  {Hentschel}}, \ and\ \bibinfo {author} {\bibfnamefont {I.}~\bibnamefont
  {Procaccia}},\ }\href@noop {} {\bibfield  {journal} {\bibinfo  {journal}
  {Phys. Rev. Lett.}\ }\textbf {\bibinfo {volume} {109}},\ \bibinfo {pages}
  {255502} (\bibinfo {year} {2012})}\BibitemShut {NoStop}%
\bibitem [{\citenamefont {Dasgupta}\ \emph {et~al.}(2013)\citenamefont
  {Dasgupta}, \citenamefont {Hentschel},\ and\ \citenamefont
  {Procaccia}}]{Dasgupta2013}%
  \BibitemOpen
  \bibfield  {author} {\bibinfo {author} {\bibfnamefont {R.}~\bibnamefont
  {Dasgupta}}, \bibinfo {author} {\bibfnamefont {H.~G.~E.}\ \bibnamefont
  {Hentschel}}, \ and\ \bibinfo {author} {\bibfnamefont {I.}~\bibnamefont
  {Procaccia}},\ }\href@noop {} {\bibfield  {journal} {\bibinfo  {journal}
  {Phys. Rev. E}\ }\textbf {\bibinfo {volume} {87}},\ \bibinfo {pages} {022810}
  (\bibinfo {year} {2013})}\BibitemShut {NoStop}%
\bibitem [{\citenamefont {Efrati}\ \emph {et~al.}(2013)\citenamefont {Efrati},
  \citenamefont {Sharon},\ and\ \citenamefont {Kupferman}}]{Efrati2013}%
  \BibitemOpen
  \bibfield  {author} {\bibinfo {author} {\bibfnamefont {E.}~\bibnamefont
  {Efrati}}, \bibinfo {author} {\bibfnamefont {E.}~\bibnamefont {Sharon}}, \
  and\ \bibinfo {author} {\bibfnamefont {R.}~\bibnamefont {Kupferman}},\
  }\href@noop {} {\bibfield  {journal} {\bibinfo  {journal} {Soft Matter}\
  }\textbf {\bibinfo {volume} {9}},\ \bibinfo {pages} {8187} (\bibinfo {year}
  {2013})}\BibitemShut {NoStop}%
\end{thebibliography}%

		\appendix

	\end{document}